\documentclass[12pt,thmsa]{article}
\usepackage{amsfonts}

\usepackage{sw20aip}

\input{tcilatex}
\begin{document}

\author{Ron Rubin \\
%EndAName
Department of Mathematics\\
Massachusetts Institute of Technology\\
and the University of the Middle East Project\\
Email: rubin@math.mit.edu \and Nathan Salwen \\
%EndAName
Lyman Laboratory of Physics\\
Harvard University\\
Email: salwen@fas.harvard.edu}
\title{A Canonical Quantization of the Baker's Map\thanks{%
30 pages, 2 figures, 1 table}}
\date{June 22, 1998}
\maketitle

\begin{abstract}
We present here a canonical quantization for the baker's map. The method we
use is quite different from that used in Balazs and Voros (ref. \cite{BV})
and Saraceno (ref. \cite{S}). We first construct a natural ``baker covering
map'' on the plane $\mathbb{R}^{2}$. We then use as the quantum algebra of
observables the subalgebra of operators on $L^{2}\left( \mathbb{R}\right) $
generated by $\left\{ \exp \left( 2\pi i\widehat{x}\right) ,\exp \left( 2\pi
i\widehat{p}\right) \right\} $ . We construct a unitary propagator such that
as $\hbar \rightarrow 0$ the classical dynamics is returned. For Planck's
constant $h=1/N$, we show that the dynamics can be reduced to the dynamics
on an $N$-dimensional Hilbert space, and the unitary $N\times N$ matrix
propagator is the same as given in ref. \cite{BV} except for a small
correction of order $h$. This correction is shown to preserve the classical
symmetry $x\rightarrow 1-x$ and $p\rightarrow 1-p$ in the quantum dynamics
for periodic boundary conditions.
\end{abstract}

\section{Introduction}

\smallskip The classical baker's map is a mapping of the unit torus onto
itself defined as follows. Let $x$ and $p$ be the coordinates on the torus
and take 
\begin{equation}
(x,p)\rightarrow (x^{\prime },p^{\prime })=\left\{ 
\begin{array}{ll}
(2x,p/2), & 0\leq x<1/2; \\ 
(2x-1,p/2+1/2), & 1/2\leq x<1.
\end{array}
\right.  \label{Classical baker's map}
\end{equation}

This map describes a stretching in $x$, shrinking in $p$, and chopping and
stacking, similar to the way bakers make certain pastries. (See Figure 1.)
The motion on the torus is completely chaotic with a positive Liapunov
exponent $\log 2$, and is in fact a paradigm for the study of classical
chaos. For more details on the classical baker's map, including a
description of the map as a dynamics on binary digits, we refer the reader
to ref. \cite{BV}.

A quantum version of the map was introduced by Balazs and Voros (ref. \cite
{BV}) and then by Saraceno (ref. \cite{S}). In this description, the
dynamics is quantized for values of Planck's constant satisfying $h=1/N$, by
constructing a quantum propagator $u_{mn}$ as a unitary $N\times N$ matrix.
The classical limit is demonstrated numerically for $N\rightarrow \infty $.
We present here a natural quantum propagator which, except for a correction
of order $\hbar $, ``reduces'' to the finite dimensional matrix propagator
given in ref. \cite{BV} at the point $\theta =\left( 0,0\right) $,
corresponding to periodic boundary conditions of the quantum wave vectors
defined below. This correction is what preserves the classical symmetry
broken in the quantization scheme presented in ref. \cite{BV}. Whether other
points on the $\theta $-torus, including the anti-periodic case studied by
Saraceno (ref. \cite{S}), are invariant under the baker's map propagator
derived below will be addressed in forthcoming papers (see ref. \cite{RS}).

\section{Baker's Map on the Plane}

Our quantization procedure uses a baker covering map we construct as
folllows. The baker's map has a natural lift to the universal covering $%
\mathbb{R}^{2}\;$of $\mathbb{T}^{2}$ given by 
\begin{equation}
\beta :(x,p)\rightarrow (x^{\prime },p^{\prime })=\left\{ 
\begin{array}{ll}
(2x,p/2), & (x,p)\in l\cap e_{p}; \\ 
(2x-1,p/2+1/2), & (x,p)\in r\cap e_{p}; \\ 
(2x+1,p/2+1/2), & (x,p)\in l\cap o_{p}; \\ 
(2x,p/2), & (x,p)\in r\cap o_{p},
\end{array}
\right.  \label{coveringmap}
\end{equation}
so that for $a,b\in \mathbb{Z}$, 
\begin{equation}
\beta ^{*}e^{2\pi i\left( ax+bp\right) }=e^{4\pi iax}e^{i\pi bp}\left( \chi
_{l}\left( x\right) +\left( -1\right) ^{b}\chi _{r}\left( x\right) \right)
\left( \chi _{e_{p}}\left( p\right) +\left( -1\right) ^{b}\chi
_{o_{p}}\left( p\right) \right) ,  \label{hope}
\end{equation}
where 
\begin{eqnarray}
l &:&=\left\langle [0,1/2)+\mathbb{Z}\right\rangle \times \mathbb{R}, 
\nonumber \\
r &:&=\left\langle [1/2,1)+\mathbb{Z}\right\rangle \times \mathbb{R},
\label{regone} \\
e_{p} &:&=\mathbb{R}\mathbf{\times }\left\langle [0,1)+2\mathbb{Z}%
\right\rangle ,  \nonumber \\
o_{p} &:&=\mathbb{R}\mathbf{\times }\left\langle [1,2)+2\mathbb{Z}%
\right\rangle .  \nonumber
\end{eqnarray}
and 
\begin{eqnarray*}
\chi _{l}(x) &=&\left\{ 
\begin{array}{ll}
1, & x\in [0,1/2)+\mathbf{Z} \\ 
0, & otherwise
\end{array}
\right. , \\
\chi _{r}(x) &=&\left\{ 
\begin{array}{ll}
1, & x\in [1/2,1)+\mathbf{Z} \\ 
0, & otherwise
\end{array}
\right. , \\
\chi _{e_{p}}(p) &=&\left\{ 
\begin{array}{ll}
1, & p\in [0,1)+2\mathbf{Z} \\ 
0, & otherwise
\end{array}
\right. , \\
\chi _{o_{p}}(p) &=&\left\{ 
\begin{array}{ll}
1, & p\in [1,2)+2\mathbf{Z} \\ 
0, & otherwise
\end{array}
\right. .
\end{eqnarray*}
We can see what this covering map does to the plane in Figure 2. There are
two important observations to make regarding this figure. The first is that
every ``LEFT'' region gets mapped into a ``BOTTOM'' region and every
``RIGHT'' into a ``TOP.'' Thus when the dynamics is given modulo $1$, we
return precisely the baker's map on the torus. The second is that there are,
of course, many such natural covering maps. Another, for instance, would be
to transport each fundamental domain to the first square ($[0,1)\times [0,1)$%
), perform the baker transformation (eqn. \ref{Classical baker's map}), then
shift back. The problem is this procedure requires a different algebraic
form for the map in each fundamental domain of the plain. The benefit of the
baker covering map presented here is there are only four fundamental regions
required to write the map.

We can also write down the inverse of the baker covering map: 
\begin{equation}
\beta ^{-1}:(x,p)\rightarrow (x^{\prime },p^{\prime })=\left\{ 
\begin{array}{ll}
(x/2,2p), & (x,p)\in e_{x}\cap b; \\ 
(x/2-1/2,2p-1), & (x,p)\in o_{x}\cap b; \\ 
(x/2+1/2,2p-1), & (x,p)\in e_{x}\cap t; \\ 
(x/2,2p), & (x,p)\in o_{x}\cap t,
\end{array}
\right.  \label{inversecoveringmap}
\end{equation}
where we have used the ``conjugate'' regions 
\begin{eqnarray}
b &:&=\mathbb{R}\mathbf{\times }\left\langle [0,1/2)+\mathbb{Z}\right\rangle
,  \nonumber \\
t &:&=\mathbb{R}\mathbf{\times }\left\langle [1/2,1)+\mathbb{Z}\right\rangle
,  \label{regtwo} \\
e_{x} &:&=\left\langle [0,1)+2\mathbb{Z}\right\rangle \mathbf{\times }%
\mathbb{R},  \nonumber \\
o_{x} &:&=\left\langle [1,2)+2\mathbb{Z}\right\rangle \times \mathbb{R}. 
\nonumber
\end{eqnarray}
Observe that these subsets of $\mathbb{R}^{2}$ satisfy the following
relations: 
\begin{eqnarray*}
l\cup r &=&b\cup t=e_{p}\cup o_{p}=e_{x}\cup o_{x}=\mathbb{R}^{2}, \\
l\cap r &=&b\cap t=e_{p}\cap o_{p}=e_{x}\cap o_{x}=\mathbf{\emptyset .}
\end{eqnarray*}

\section{The Quantum Propagator}

Our quantization can now be described. We follow the canonical procedure of
mapping the function $x$ to the multiplication operator $\widehat{x}$ and
the function $p$ to the derivative operator $\widehat{p}=\left( \hbar
/i\right) d/dx$. We call this mapping $Q_{\hbar }$ and note that it is
non-unique due to ordering ambiguities. For the case of the torus, this
procedure is used commonly in the mathematics literature (see refs. \cite
{LRS}, \cite{KLMR}, \cite{E}, \cite{DEG}, \cite{Z}), but much less so in
physics papers. The quantum algebra of observables $\frak{A}_{\hbar }$ is
restricted to the set operators generated by the quantization of the
classical generators: $Q_{\hbar }\left( \exp \left( 2\pi ix\right) \right)
=\exp \left( 2\pi i\widehat{x}\right) $ and $Q_{\hbar }\left( \exp \left(
2\pi ip\right) \right) =\exp \left( 2\pi i\widehat{p}\right) $. Following
the standard notation, we let 
\begin{eqnarray*}
U &=&\exp \left( 2\pi i\widehat{x}\right) \text{,} \\
V &=&\exp \left( 2\pi i\widehat{p}\right)
\end{eqnarray*}
so that 
\[
UV=e^{4\pi ^{2}i\hbar }VU\text{.} 
\]
We construct a quantum propagator by quantizing the dynamics of the covering
map eqn. \ref{coveringmap}. The quantum dynamics induced on the algebra of
observables for the quantum torus is the quantum baker's map.

We now construct the quantum propagator $F$. The kinematics is already
given: the Hilbert space is the usual $L^{2}(\mathbb{R})$. For the dynamics,
we work in the Heisenberg picture, and first give the quantum analogs of
eqns. \ref{regone} and \ref{regtwo}. We define the following projection
operators: 
\begin{eqnarray}
L &:&=\int_{[0,1/2)+\mathbb{Z}}\left| x\right\rangle \left\langle x\right|
\;dx,  \label{Projone} \\
R &:&=\int_{[1/2,1)+\mathbb{Z}}\left| x\right\rangle \left\langle x\right|
\;dx,  \nonumber \\
B &:&=\int_{[0,1/2)+\mathbb{Z}}\left| p\right\rangle \left\langle p\right|
\;dp,  \nonumber \\
T &:&=\int_{[1/2,1)+\mathbb{Z}}\left| p\right\rangle \left\langle p\right|
\;dp,  \nonumber
\end{eqnarray}
and 
\begin{eqnarray}
E_{x} &:&=\int_{[0,1)+2\mathbb{Z}}\left| x\right\rangle \left\langle
x\right| \;dx,  \label{Projtwo} \\
O_{x} &:&=\int_{[1,2)+2\mathbb{Z}}\left| x\right\rangle \left\langle
x\right| \;dx,  \nonumber \\
E_{p} &:&=\int_{[0,1)+2\mathbb{Z}}\left| p\right\rangle \left\langle
p\right| \;dp,  \nonumber \\
O_{p} &:&=\int_{[1,2)+2\mathbb{Z}}\left| p\right\rangle \left\langle
p\right| \;dp.  \nonumber
\end{eqnarray}
Observe that 
\[
L+R=B+T=E_{x}+O_{x}=E_{p}+O_{p}=I, 
\]
and 
\[
LR=BT=E_{x}O_{x}=E_{p}O_{p}=0. 
\]
We next define appropriate ``shift'' operators. A shift in $p,$ or a shift
in $x$, by unity is achieved by the following unitary operators,
respectively: 
\begin{eqnarray*}
X &=&e^{i\widehat{x}/\hbar }, \\
Y &=&e^{i\widehat{p}/\hbar }.
\end{eqnarray*}
Note that $X$ and $Y$ commute with the algebra $\frak{A}_{\hbar }$ generated
by $U$ and $V$: 
\[
\left[ X,U\right] =\left[ Y,U\right] =\left[ X,V\right] =\left[ Y,V\right] =0%
\text{.} 
\]
We shall also need the following commutation relations: 
\begin{eqnarray}
XL &=&LX,\quad YL=LY,\quad XR=RX,\quad YR=RY,  \label{commutation relations}
\\
Y^{1/2}L &=&RY^{1/2},\quad X^{1/2}B=TX^{1/2},  \nonumber \\
YE_{x} &=&O_{x}Y,\quad XE_{p}=O_{p}X.  \nonumber
\end{eqnarray}
We demonstrate one of these commutation relations explicitly. The others
involve similar calculations: 
\begin{eqnarray*}
Y^{1/2}L &=&e^{i\widehat{p}/2\hbar }\int_{[0,1/2)+\mathbb{Z}}\left|
x\right\rangle \left\langle x\right| \;dxe^{-i\widehat{p}/2\hbar }e^{i%
\widehat{p}/2\hbar } \\
&=&\int_{[0,1/2)+\mathbb{Z}}\left| x-1/2\right\rangle \left\langle
x-1/2\right| \;dxe^{i\widehat{p}/2\hbar } \\
&=&\int_{[1/2,1)+\mathbb{Z}}\left| x\right\rangle \left\langle x\right|
\;dxe^{i\widehat{p}/2\hbar }=RY^{1/2}.
\end{eqnarray*}

We next find the unitary operator $S$ which takes $\widehat{x}$ to $2%
\widehat{x}$ and $\widehat{p}$ to $\widehat{p}/2$. Formally, for any
operator expandable as a Taylor series in $\widehat{x}\widehat{p}+\widehat{p}%
\widehat{x}$, we find 
\[
\widehat{x}f\left( \widehat{x}\widehat{p}+\widehat{p}\widehat{x}\right)
=f\left( \widehat{x}\widehat{p}+\widehat{p}\widehat{x}+2i\hbar \right) 
\widehat{x}. 
\]
Thus, if we define the operator 
\begin{equation}
S:={\large \exp }\left( -\frac{i\log 2}{2\hbar }(\widehat{x}\widehat{p}+%
\widehat{p}\widehat{x})\right)  \label{stretching}
\end{equation}
we see 
\begin{eqnarray*}
\widehat{x}S &=&2S\widehat{x}, \\
\widehat{p}S &=&S\widehat{p}/2,
\end{eqnarray*}
or 
\begin{eqnarray*}
S^{\dagger }\widehat{x}S &=&2\widehat{x}, \\
S^{\dagger }\widehat{p}S &=&\widehat{p}/2.
\end{eqnarray*}
Observe also that the operator $S$ is unitary since $\widehat{x}\widehat{p}+%
\widehat{p}\widehat{x}$ is Hermitian.

We are now in a position to write down a propagator for the baker's map.
Based on eqn. \ref{hope}, we have the following definition.

\begin{definition}
\textbf{(Baker's Map Propagator)} Let the operator $F$ be defined as
follows: 
\begin{eqnarray}
F &=&S(L+X^{-1}R)(E_{p}+Y^{-1/2}O_{p})  \label{propagator} \\
&=&(E_{x}+X^{-1/2}O_{x})(B+Y^{-1}T)S  \nonumber
\end{eqnarray}
\end{definition}

\begin{lemma}
$F$ is unitary.
\end{lemma}

%TCIMACRO{\TeXButton{Proof}{\proof}}
%BeginExpansion
\proof%
%EndExpansion
Observe that $E_{x}=E_{x}^{\dagger }$, $O_{x}=O_{x}^{\dagger }$, $%
B=B^{\dagger }$, $T=T^{\dagger }$. It follows that 
\begin{eqnarray*}
(L+X^{-1}R)(L+X^{-1}R)^{\dagger } &=&L+R=I, \\
(E_{p}+Y^{-1/2}O_{p})(E_{p}+Y^{-1/2}O_{p})^{\dagger } &=&E_{p}+O_{p}=I.
\end{eqnarray*}
Thus $F$ is the product of three unitary operators, hence is unitary. 
%TCIMACRO{\TeXButton{End Proof}{\endproof}}
%BeginExpansion
\endproof%
%EndExpansion

\section{The Classical Limit}

Because of the discontinuity of the baker's map (and its covering map), the
classical limit requires more thought (see, for instance, ref. \cite{DEG}).
The basic problem comes from the fact that the projection operators $L$ and $%
E_{p}$ (for example) do not commute as $\hbar \rightarrow 0$ (even weakly).
This comes from scaling - each term contributes less, but the number of
terms increases.

We demonstrate this explicitly with the following example. Consider a
function $\phi \in L^{2}\left( \mathbb{R}\right) $ which is supported only
for $x\in $ $[1/2,1)$, and a $\psi $ supported only in $[0,1/2)$. Then by
construction, we see 
\[
\left\langle \psi \right| E_{p}L\left| \phi \right\rangle =0\text{.} 
\]
Now consider the action of $LE_{p}$. We see 
\begin{eqnarray*}
\left\langle \psi \right| LE_{p}\left| \phi \right\rangle &=&\int
dxdx^{\prime }\overline{\psi \left( x\right) }\phi \left( x^{\prime }\right)
\chi _{l}\left( x\right) \left\langle x\right| E_{p}\left| x^{\prime
}\right\rangle \\
&=&\int dxdx^{\prime }\overline{\psi \left( x\right) }\phi \left( x^{\prime
}\right) \chi _{l}\left( x\right) \left\langle x\right| 1/2-(i/\pi )\sum_{k\
odd}\frac{e^{\pi ik\widehat{p}}}{k}\left| x^{\prime }\right\rangle \\
&=&\frac{1}{2}\int dxdx^{\prime }\chi _{l}\left( x\right) \overline{\psi
\left( x\right) }\phi \left( x\right) \\
&&-\frac{i}{\pi }\int dxdx^{\prime }\chi _{l}\left( x\right) \overline{\psi
\left( x\right) }\phi \left( x^{\prime }\right) \left\langle x\right|
\sum_{k\ odd}\frac{e^{\pi ik\widehat{p}}}{k}\left| x^{\prime }\right\rangle .
\end{eqnarray*}
The first term in the expression above is zero, while the second term yields 
\begin{eqnarray*}
\left\langle \psi \right| LE_{p}\left| \phi \right\rangle &=&-\frac{i}{\pi }%
\sum_{k\text{ odd}}\frac{1}{k}\int dxdx^{\prime }\chi _{l}\left( x\right) 
\overline{\psi \left( x\right) }\phi \left( x^{\prime }\right) \left\langle
x\right| \left. x^{\prime }-\pi k\hbar \right\rangle \\
&=&-\frac{i}{\pi }\sum_{k\text{ odd}}\frac{1}{k}\int dx\chi _{l}\left(
x\right) \overline{\psi \left( x\right) }\phi \left( x+\pi \hbar k\right) .
\end{eqnarray*}

As a particular example, we choose 
\begin{eqnarray*}
\phi \left( x\right) &=&\chi _{[1/2,1)}\left( x\right) , \\
\psi \left( x\right) &=&\chi _{[0,1/2)}\left( x\right) .
\end{eqnarray*}
Then 
\begin{eqnarray*}
\left\langle \psi \right| LE_{p}\left| \phi \right\rangle &=&-\frac{i}{\pi }%
\sum_{k\text{ odd}}\frac{1}{k}\int_{0}^{1/2}dx\chi _{[1/2,1)}\left( x+\pi
\hbar k\right) \\
&=&-\frac{i}{\pi }\sum_{k\text{ odd}}\frac{1}{k}\int_{0}^{1/2}dx\chi
_{[1/2-\pi \hbar k,1-\pi \hbar k)}\left( x\right)
\end{eqnarray*}
The only terms in the sum not equal to zero have $0\leq k<\frac{1}{2\pi
\hbar }$ or $\frac{1}{2\pi \hbar }\leq k<\frac{1}{\pi \hbar }$, so we see 
\begin{eqnarray*}
\left\langle \psi \right| LE_{p}\left| \phi \right\rangle &=&-\frac{i}{\pi }%
\sum\Sb k\text{ odd}  \\ \frac{1}{2}\leq \pi \hbar k<1  \endSb \frac{1}{k}%
\left( 1-\pi \hbar k\right) -\frac{i}{\pi }\sum\Sb k\text{ odd}  \\ 0\leq
\pi \hbar k<1/2  \endSb \frac{1}{k}\left( \pi \hbar k\right) \\
&\rightarrow &-\frac{i}{\pi }\left( \int_{1/2\pi \hbar }^{1/\pi \hbar
}\left( \frac{1}{k}-\pi \hbar \right) dk+\int_{0}^{1/2\pi \hbar }\pi \hbar
\,dk\right) \\
&=&-\frac{i}{\pi }\left( \left( \log 2-1/2\right) +1/2\right) \\
&=&-\frac{i\log 2}{\pi }\neq 0.
\end{eqnarray*}
Thus the limit is finite but nonzero.

All is not lost, however, as we can take a more constrained view of what
constitutes a quantum state with a classical limit (see, for example, ref. 
\cite{Hep}). We take as our quantum state the coherent state $\left| \phi
_{\hbar };x_{0},p_{0}\right\rangle $ centered around the point $\left(
x_{0},p_{0}\right) $ with a width of $\sqrt{\hbar }$. Recall that a coherent
state can be written 
\begin{equation}
\phi _{x_{0},p_{0}}^{\hbar }\left( x\right) =\left\langle x\right. \left|
\phi _{\hbar };x_{0},p_{0}\right\rangle =\frac{1}{\left( \pi \hbar \right)
^{1/4}}e^{-\left( x-x_{0}\right) ^{2}/2\hbar }e^{ip_{0}x/\hbar
-ip_{0}x_{0}/2\hbar }  \label{coherent state}
\end{equation}
with a Fourier transform 
\[
\widetilde{\phi }_{x_{0},p_{0}}^{\hbar }\left( p\right) =\left\langle
p\right. \left| \phi _{\hbar };x_{0},p_{0}\right\rangle =\frac{1}{\left( \pi
\hbar \right) ^{1/4}}e^{-\left( p-p_{0}\right) ^{2}/2\hbar
}e^{-ipx_{0}/\hbar +ip_{0}x_{0}/2\hbar }. 
\]

We can define the classical limit in terms of these coherent states.

\begin{definition}
A quantum propagator $F$ is said to have a weak classical limit if for any
function on the torus $A$, and for almost every $x_{0},p_{0}$, 
\[
\lim_{\hbar }\left\langle \phi _{\hbar };x_{0},p_{0}\right| F^{\dagger
}Q_{\hbar }\left( A\right) F\left| \phi _{\hbar };x_{0},p_{0}\right\rangle
-\left\langle \phi _{\hbar };\beta \left( x_{0},p_{0}\right) \right|
Q_{\hbar }\left( A\right) \left| \phi _{\hbar };\beta \left(
x_{0},p_{0}\right) \right\rangle 
\]
where $\beta $ is the classical evolution.
\end{definition}

\smallskip In other words, if, as $\hbar \rightarrow 0$, any observable has
the same value under classical and quantum evolution for almost all wave
packets, we say the quantum mechanics yields the classical mechanics. Note
also that this differs from the definition given in ref. \cite{Hep} by use
of the ``almost all'' caveat: we allow the classical limit to fail at a set
of measure zero points in phase space.

We then have the following theorem.

\begin{theorem}
The propagator $F$ defined in eqn. \ref{propagator} has a weak classical
limit in the sense of definition $3$.
\end{theorem}

Proof. We give the proof for a harmonic $Q_{\hbar }\left( A\right)
=U^{a}V^{b}$. The general case will follow by linearity and continuity. We
divide the proof into steps.

Step 1. We first calculate the expectation value of the operator $U^{a}V^{b}$
in the coherent states. We see 
\begin{eqnarray*}
&&\left\langle \phi _{\hbar };x_{0},p_{0}\right| U^{a}V^{b}\left| \phi
_{\hbar };x_{0},p_{0}\right\rangle \\
&=&\frac{1}{\left( \pi \hbar \right) ^{1/2}}\int dxdx^{\prime }e^{-\left(
x-x_{0}\right) ^{2}/2\hbar }e^{-ip_{0}x/\hbar }e^{2\pi iax}\delta \left(
x-x^{\prime }-2\pi b\hbar \right) e^{-\left( x^{\prime }-x_{0}\right)
^{2}/2\hbar }e^{ip_{0}x^{\prime }/\hbar } \\
&=&e^{2\pi ibp_{0}}e^{2\pi iax_{0}}e^{-2\pi ^{2}b^{2}\hbar }e^{-4\pi
^{2}\hbar \left( b+ia\right) ^{2}}\rightarrow e^{2\pi ibp_{0}}e^{2\pi
iax_{0}}\quad \text{as}\quad \hbar \rightarrow 0\text{.}
\end{eqnarray*}
Step 2. Observe that acting on these states, we see 
\begin{eqnarray*}
\left\| L\left| \phi _{\hbar };x_{0},p_{0}\right\rangle \right\| ^{2}
&=&\left\langle \phi _{\hbar };x_{0},p_{0}\right| L\left| \phi _{\hbar
};x_{0},p_{0}\right\rangle \\
&=&\frac{1}{\left( \pi \hbar \right) ^{1/2}}\sum_{k\in \mathbb{Z}%
}\int_{0}^{1/2}e^{-\left( x+k-x_{0}\right) ^{2}/\hbar }dx.
\end{eqnarray*}
Suppose $x_{0}\neq l/2$ for $l\in \mathbb{Z}$. Now choose $\epsilon >0$ and $%
l\in \mathbb{Z}$ such that $\left. x_{0}\in \left[ l/2+\epsilon ,\left(
l+1\right) /2-\epsilon \right) \right. $ for $\epsilon >0$.

For the case of $l$ odd, we see that the value of the integral is bounded by 
\[
\left| \frac{1}{\left( \pi \hbar \right) ^{1/2}}\sum_{k\in \mathbb{Z}%
}\int_{0}^{1/2}e^{-\left( x-x_{0}+k\right) ^{2}/\hbar }\,dx\right| \leq 
\frac{2}{\sqrt{\pi }}\int_{\epsilon /\sqrt{\hbar }}^{\infty }e^{-x^{2}}dx. 
\]
A bound on this integral can easily be given for $\epsilon /\sqrt{\hbar }>1.$
We see that 
\[
\int_{\epsilon /\sqrt{\hbar }}^{\infty }e^{-x^{2}}dx\leq \int_{\epsilon /%
\sqrt{\hbar }}^{\infty }xe^{-x^{2}}dx=\frac{1}{2}\int_{\epsilon /\sqrt{\hbar 
}}^{\infty }e^{-u}du=e^{-\epsilon ^{2}/\hbar }/2. 
\]
Thus for $l$ odd, the limit of the integral is zero as $\hbar \rightarrow 0$.

Now consider $x_{0}\in [l/2+\epsilon ,\left( l+1\right) /2-\epsilon )$ with $%
l$ even. Then we see that 
\begin{eqnarray*}
\left\| L\left| \phi _{\hbar };x_{0},p_{0}\right\rangle \right\| ^{2}
&=&\left\| \left( I-R\right) \left| \phi _{\hbar };x_{0},p_{0}\right\rangle
\right\| ^{2} \\
&=&1-\left\| R\left| \phi _{\hbar };x_{0},p_{0},\mu \right\rangle \right\|
^{2} \\
&\rightarrow &1\quad \text{as\ }\hbar \rightarrow 0\text{.}
\end{eqnarray*}
Similar results hold for all the projection operators $%
L,R,B,T,E_{x},O_{x},E_{p},O_{p}$ defined in eqns. \ref{Projone} and \ref
{Projtwo}. We let $\tilde{l},\tilde{r},\tilde{b},\tilde{t},\tilde{e}_{x},%
\tilde{o}_{x},\tilde{e}_{p},\tilde{o}_{p}$ denote the interior of the
regions given in eqns. \ref{regone} and \ref{regtwo}, that is the regions
with the boundaries removed. (For instance $\widetilde{l}$ does not contain $%
x=0$ or $x=1/2$.) Note that $\left( \widetilde{l}\cap \widetilde{e}%
_{p}\right) \cup \left( \widetilde{r}\cap \widetilde{e}_{p}\right) \cup
\left( \widetilde{l}\cap \widetilde{o}_{p}\right) \cup \left( \widetilde{r}%
\cap \widetilde{o}_{p}\right) $ is dense in $\mathbb{R}^{2}$. We summarize
these results in Table 1.

Step 3. Now suppose $\left( x_{0},p_{0}\right) \in \tilde{r}\cap \widetilde{o%
}_{p}$. Then consider the quantum evolution. We see 
\begin{eqnarray*}
&&\left\langle \phi _{\hbar };x_{0},p_{0}\right| \left(
E_{p}+Y^{1/2}O_{p}\right) \left( L+XR\right) S^{\dagger }U^{a}V^{b}S\left(
L+X^{-1}R\right) \left( E_{p}+Y^{-1/2}O_{p}\right) \left| \phi _{\hbar
};x_{0},p_{0}\right\rangle \\
&=&\left\langle \phi _{\hbar };x_{0},p_{0}\right| \left(
E_{p}+Y^{1/2}O_{p}\right) \left( L+XR\right) U^{2a}V^{b/2}\left(
L+X^{-1}R\right) \left( E_{p}+Y^{-1/2}O_{p}\right) \left| \phi _{\hbar
};x_{0},p_{0}\right\rangle
\end{eqnarray*}
Multiplying out, we see $16$ terms in the expansion. Consider one of these
terms. We see from Table 1 that 
\begin{eqnarray*}
&&\left| \left\langle \phi _{\hbar };x_{0},p_{0}\right|
E_{p}XRU^{2a}V^{b/2}LY^{-1/2}O_{p}\left| \phi _{\hbar
};x_{0},p_{0}\right\rangle \right| \\
&=&\left| \left\langle \phi _{\hbar };x_{0},p_{0}\right|
E_{p}RU^{2a+N}V^{\left( b-N\right) /2}RO_{p}\left| \phi _{\hbar
};x_{0},p_{0}\right\rangle \right| \\
&\leq &\left| \left\langle \phi _{\hbar };x_{0},p_{0}\right| E_{p}\left|
\phi _{\hbar };x_{0},p_{0}\right\rangle \right| \\
&&\times \left| \left\langle \phi _{\hbar };x_{0},p_{0}\right| \left(
RU^{2a+N}V^{\left( b-N\right) /2}RO_{p}\right) ^{\dagger }\left(
RU^{2a+N}V^{\left( b-N\right) /2}RO_{p}\right) \left| \phi _{\hbar
};x_{0},p_{0}\right\rangle \right| \\
&\rightarrow &0\text{\quad as\quad }\hbar \rightarrow 0\text{\quad since }%
\left( x_{0},p_{0}\right) \in \widetilde{o}_{p}\text{.}
\end{eqnarray*}
Note in the second step we have used the Schwartz inequality and the
identity $E_{p}^{2}=E_{p}$. Similarly, $15$ of the terms vanish as $\hbar
\rightarrow 0$. The only surviving term for $\left( x_{0},p_{0}\right) \in 
\tilde{r}\cap \widetilde{o}_{p}$ is 
\begin{eqnarray*}
&&\left\langle \phi _{\hbar };x_{0},p_{0}\right|
O_{p}Y^{1/2}LU^{2a}V^{b/2}LY^{-1/2}O_{p}\left| \phi _{\hbar
};x_{0},p_{0}\right\rangle \\
&\rightarrow &\left\langle \phi _{\hbar };x_{0},p_{0}\right|
U^{2a}V^{b/2}\left| \phi _{\hbar };x_{0},p_{0}\right\rangle \quad \\
&\rightarrow &e^{2\pi i\left( 2ax_{0}+\left( b/2\right) p_{0}\right) }.
\end{eqnarray*}
Step 4. Now consider the classical evolution. We have, for $\left(
x_{0},p_{0}\right) \in \tilde{r}\cap \widetilde{o}_{p}$%
\begin{eqnarray*}
\left\langle \phi _{\hbar };\beta \left( x_{0},p_{0}\right) \right|
U^{a}V^{b}\left| \phi _{\hbar };\beta \left( x_{0},p_{0}\right)
\right\rangle &=&\left\langle \phi _{\hbar };2x_{0},p_{0}/2\right|
U^{a}V^{b}\left| \phi _{\hbar };2x_{0},p_{0}/2\right\rangle \\
&\rightarrow &e^{2\pi i\left( 2ax_{0}+\left( b/2\right) p_{0}\right) }\text{.%
}
\end{eqnarray*}
The calculations for the other regions are similar, and we omit the details
here.This concludes the proof. 
%TCIMACRO{\TeXButton{End Proof}{\endproof}}
%BeginExpansion
\endproof%
%EndExpansion

\section{\protect\smallskip Planck's Constant $=1/N$}

A remarkable set of properties can be associated with quantum dynamics on
the torus if we let Planck's constant satisfy the integrality condition 
\[
h=1/N. 
\]
This fact is evidenced by the quantization schemes presented in ref. \cite
{BV} and ref. \cite{S} for the baker's map. In ref. \cite{KLMR}, a procedure
was developed to obtain a finite-dimensional quantum cat-dynamics for the
case of periodic boundary conditions, and it was found to have the same form
as the original matrix quantization proposed in ref. \cite{HB}.

We shall find for the baker's map, a finite-dimensional matrix propagator
for $N$ even given by 
\begin{equation}
\left( \Phi _{n}^{\left( 0,0\right) },F\Phi _{m}^{\left( 0,0\right) }\right)
_{P}=\left[ \left( \mathcal{Z}\right) \left( \mathcal{F}^{N}\right)
^{-1}\left( 
\begin{array}{ll}
\mathcal{F}^{N/2} & 0 \\ 
0 & -\mathcal{F}^{N/2}
\end{array}
\right) \left( \mathcal{Z}^{-2}\right) \right] _{nm}  \label{OURS}
\end{equation}
where the indices $n$ and $m$ take values between $0$ and $N-1$ and we use
the following notations: (1) $\mathcal{F}^{N}$ is the $N\times N$ discrete
Fourier transform matrix defined in eqn. \ref{Fourier Transform} below; (2) $%
\mathcal{Z}$ is the diagonal matrix 
\begin{equation}
\left( \mathcal{Z}\right) _{nm}=\delta _{n,m}e^{i\pi n/N};  \label{phase}
\end{equation}
(3) $\Phi _{n}^{\left( 0,0\right) }$ is a basis vector of the Hilbert space $%
\mathcal{H}_{\hbar }\left( 0\right) \cong \Bbb{C}^{N}$ defined as the
periodic $\delta $-comb (eqn. \ref{delta comb}); and (4) $\left( \cdot
,\cdot \right) _{P}$ is the inner product defined in eqn. \ref{ip}.

Since $\Phi _{n+N}^{\left( 0,0\right) }=\Phi _{n}^{\left( 0,0\right) }$, the
extended matrix for general $n$ and $m\in \mathbb{Z}$ is periodic with
period $N$. Thus the form of eqn. \ref{phase} may be misleading for $n,m$
outside of the fundamental range $\left[ 0,N-1\right] $. In general, we
write $\left( \mathcal{Z}\right) _{nm}=\delta _{n,m}e^{i\pi \left(
n/N-\left[ n/N\right] \right) }$ where $\left[ n/N\right] $ represents the
integer part of $n/N$.

In ref. \cite{BV}, the following matrix propagator was proposed by Balazs
and Voros: 
\begin{equation}
\sum_{a=0}^{N/2-1}\left( \mathcal{F}^{N}\right) _{na}^{-1}\left( 
\begin{array}{ll}
\mathcal{F}^{N/2} & 0 \\ 
0 & \mathcal{F}^{N/2}
\end{array}
\right) _{am}  \label{Balazs Voros}
\end{equation}
It can be easily verified that for $n$ even, the results are identical,
while for $n$ odd, this differs from eqn. \ref{OURS} by a phase $e^{i\pi
\zeta /N}$ where $\zeta =n-2m\,$ for $0\leq m<N/2$ or $\zeta =n-2\left(
m-N/2\right) $ for $N/2\leq m<N$. These phases become small along classical
trajectories.

Indeed, in the small $\hbar $ $\,$limit we can see that these phases are $%
O\left( \hbar \right) $ corrections. Consider, for example the case $0\leq
m<N/2$. In \cite{BV}, it was shown that the matrix elements (eqn. \ref
{Balazs Voros}) for $n$ odd can be written as 
\[
\frac{\sqrt{2}}{N}\left( 1+i\cot \frac{\pi }{N}\left( n-2m\right) \right) 
\text{.} 
\]
We let $n=N\chi $ and $m=N\xi $ with $\chi ,\xi \in \left[ 0,1/2\right] $ to
see the behavior as $N\rightarrow \infty $. Then, the above expression
behaves like 
\[
\frac{\sqrt{2}}{N}\left( 1+i\cot \pi \left( \chi -2\xi \right) \right)
\rightarrow 0 
\]
unless $\chi -2\xi $ $=O\left( 1/N\right) =O\left( \hbar \right) $. Thus the
phase change is $O\left( \hbar \right) $ for the non-vanishing matrix
elements.

Furthermore, the matrix propagator (eqn. \ref{OURS}) is shown below to
preserve the symmetry $x\rightarrow 1-x$ and $p\rightarrow 1-p$, which is
not preserved in the original quantization given by Balazs and Voros \cite
{BV}. (In Saraceno \cite{S}, an anti-periodic quantization is formulated
which does preserve this symmetry.) In addition, we verify that this new
matrix propagator satisfies the classical symmetry $x\rightarrow p,$ $%
p\rightarrow x,$ and $t\rightarrow -t$.

\subsection{\protect\smallskip The $\theta $-torus}

For $h=1/N$ the algebra generated by $U$ and $V$ has a natural center
generated by 
\begin{eqnarray}
X &=&U^{N},  \label{center} \\
Y &=&V^{N}.  \nonumber
\end{eqnarray}
That is 
\[
\left[ X,Y\right] =\left[ X,U\right] =\left[ X,V\right] =\left[ Y,U\right]
=\left[ Y,V\right] =0. 
\]
In ref. \cite{KLMR} and \cite{RL}, this insight was used to show that $%
L^{2}\left( \mathbb{R}\right) $ can be decomposed via the following
eigenvalue problem: 
\begin{eqnarray*}
X\Phi &=&e^{2\pi i\theta _{1}}\Phi , \\
Y\Phi &=&e^{2\pi i\theta _{2}}\Phi ,
\end{eqnarray*}
where $\theta =\left( \theta _{1},\theta _{2}\right) \in \Bbb{T}^{2}$. As in
ref. \cite{KLMR}, let $\mathcal{H}_{\hbar }\left( \theta \right) $ denote
the space of (non-normalizable) independent eigenvectors with fixed $\theta $%
. The space $\mathcal{H}_{\hbar }\left( \theta \right) $ has a natural inner
product defined as an integral over the fundamental domain $D=[0,1]\subset %
\mathbb{R}$ given by 
\begin{equation}
\left( \Psi _{1}(\theta ),\Psi _{2}(\theta )\right) _{P}=\int_{0}^{1}%
\overline{\Psi _{1}(x,\theta )}(K\Psi _{2})(x,\theta )dx,  \label{ip}
\end{equation}
where 
\begin{eqnarray*}
K\Psi _{2}(x,\theta ) &=&\int_{-\infty }^{\infty }K\left( x,y\right) \Psi
_{2}(y,\theta )dy, \\
K\left( x,y\right) &=&\frac{1}{2\pi \hbar }g\left( \frac{x-y}{2\hbar }\right)
\end{eqnarray*}
and 
\[
g(r)=\frac{\sin r}{r}\,e^{-\hbar r^{2}+ir}. 
\]
The following lemma was proved in \cite{RL}:

\begin{lemma}
\label{bargman theta}(i) The following (generalized) functions are elements
of $\mathcal{H}_{\hbar }\left( \theta \right) $ of unit norm: 
\begin{equation}
\Phi _{m}^{(\theta )}(x)=\frac{e^{2\pi i\theta _{2}m/N}}{N^{1/2}}\sum_{k\in %
\mathbb{Z}}e^{2\pi i\theta _{2}k}\delta \left( x-\frac{m+\theta _{1}+NK}{N}%
\right) .  \label{delta comb}
\end{equation}
They are periodic in $m$, 
\begin{equation}
\Phi _{m+N}^{\left( \theta \right) }=\Phi _{m}^{\left( \theta \right) },
\label{periodicity}
\end{equation}
and furthermore, 
\[
\Phi _{0}^{\left( \theta \right) },...,\Phi _{N-1}^{\left( \theta \right) } 
\]
are orthogonal vectors in $\mathcal{H}_{\hbar }\left( \theta \right) $.

(ii) The space $\mathcal{H}_{\hbar }\left( \theta \right) $ has dimension $N$%
. Consequently the functions \ref{delta comb} form an orthonormal basis for $%
\mathcal{H}_{\hbar }\left( \theta \right) $.
\end{lemma}

We use the following notation for these vectors: 
\[
\Phi _{m}^{(\theta )}=\frac{e^{2\pi i\theta _{2}m/N}}{N^{1/2}}\sum_{k\in %
\mathbb{Z}}e^{2\pi i\theta _{2}k}\left| \frac{\theta _{1}+m}{N}%
+k\right\rangle _{x} 
\]
These are the $\delta $-comb wavefunctions seen for example in ref. \cite{HB}%
. We can, of course, just as easily work in momentum representation.

\begin{lemma}
For $h=1/N$, 
\[
\Phi _{m}^{(\theta )}=e^{-2\pi i\theta _{1}\theta _{2}/N}\sum_{n=0}^{N-1}%
\mathcal{F}_{mn}^{N}\widetilde{\Phi }_{n}^{(\theta )}, 
\]
where $\left\{ \widetilde{\Phi }_{n}^{(\theta )}\right\} _{0\leq n\leq N-1}$
are the momentum-state wave functions on the torus, 
\begin{equation}
\widetilde{\Phi }_{n}^{(\theta )}=\frac{e^{-2\pi in\theta _{1}/N}}{\sqrt{N}}%
\sum_{k}e^{-2\pi i\theta _{1}k}\left| \frac{\theta _{2}+n}{N}+k\right\rangle
_{p},  \label{p state delta comb}
\end{equation}
and $\mathcal{F}_{mn}^{N}$ is the matrix for the discrete Fourier transform, 
\begin{equation}
\mathcal{F}_{mn}^{N}=\frac{e^{-2\pi imn/N}}{\sqrt{N}}
\label{Fourier Transform}
\end{equation}
$.$
\end{lemma}

In particular, for the subsets $\theta _{1}=0$ or $\theta _{2}=0$, changing
coordinates from momentum representation to position representation is
simply a discrete Fourier transform.

\section{Dynamics at $\theta =\left( 0,0\right) $}

The point $\theta =\left( 0,0\right) $ of the $\theta $-torus corresponds to
the $N$-dimensional vector space $\mathcal{H}_{\hbar }\left( 0\right) $ of
periodic $\delta $-combs. For the quantum baker's map, we now show that $%
\theta =$ $\left( 0,0\right) $ is an invariant point of the dynamics on the $%
\theta $-torus \textit{for }$N$\textit{\ even}. That is, the set of periodic 
$\delta $-combs is mapped onto itself by our propagator $F$. We have, using
the commutation relations eqn. \ref{commutation relations}, 
\[
XF\Phi _{m}^{\left( 0,0\right) }=FX^{2}\Phi _{m}^{\left( 0,0\right) }=F\Phi
_{m}^{\left( 0,0\right) } 
\]
and 
\begin{eqnarray*}
YF\Phi _{m}^{\left( 0,0\right) } &=&\left( O_{x}+X^{-1/2}E_{x}\right) \left(
B+Y^{-1}T\right) SY^{1/2}\Phi _{m}^{\left( 0,0\right) } \\
&=&\left( X^{1/2}O_{x}+E_{x}\right) \left( T+Y^{-1}B\right)
SX^{-1}Y^{1/2}\Phi _{m}^{\left( 0,0\right) } \\
&=&\left( E_{x}+X^{1/2}O_{x}\right) \left( B+YT\right) SX^{-1}Y\Phi
_{m}^{\left( 0,0\right) } \\
&=&E_{x}\left( B+YT\right) SX^{-1}Y\Phi _{m}^{\left( 0,0\right)
}+X^{-1/2}O_{x}\left( B+YT\right) SXY\Phi _{m}^{\left( 0,0\right) } \\
&=&E_{x}\left( B+Y^{-1}T\right) S\Phi _{m}^{\left( 0,0\right)
}+X^{-1/2}O_{x}\left( B+Y^{-1}T\right) S\Phi _{m}^{\left( 0,0\right) } \\
&=&F\Phi _{m}^{\left( 0,0\right) }.
\end{eqnarray*}

Observe that we can write any $A\ $as $\sum_{j,k}\gamma _{jk}U^{j}V^{k}$.
Letting $\,j=a+Nc$ and $k=b+Nd$ with $0\leq a,b\leq N-1$ and $c,d\in \Bbb{Z}$%
, we see from eqn. \ref{center} that 
\[
U^{a+Nc}V^{b+Nd}\Phi _{m}^{\left( 0,0\right) }=U^{a}V^{b}X^{c}Y^{d}\Phi
_{m}^{\left( 0,0\right) }=U^{a}V^{b}\Phi _{m}^{\left( 0,0\right) }\text{.} 
\]
Thus, acting on the subspace $\mathcal{H}\left( 0\right) $, the algebra $%
\frak{A}_{\hbar }$ is reduced to a set of $N^{2}$ operators. This is
isomorphic to the algebra of $N\times N$ matrices. One of these operators is
the propagator $F$, and it remains to determine the matrix elements.

\begin{theorem}
The matrix elements for the propagator $F$ on the subspace $\mathcal{H}%
\left( 0\right) $ are given by eqn. \ref{OURS}.
\end{theorem}

Proof. We divide this calculation into different cases. For $0\leq m<N/2$,
we have 
\begin{eqnarray*}
F\Phi _{m}^{\left( 0,0\right) } &=&\left( E_{x}+X^{-1/2}O_{x}\right) \left(
B+Y^{-1}T\right) S\Phi _{m}^{\left( 0,0\right) } \\
&=&\left( E_{x}+X^{-1/2}O_{x}\right) \left( B+Y^{-1}T\right) \sqrt{\frac{2}{N%
}}\sum_{k\in \mathbb{Z}}\left| \frac{2m}{N}+2k\right\rangle \\
&=&\frac{1}{\sqrt{2}}\left( E_{x}+X^{-1/2}O_{x}\right) \left(
B+Y^{-1}T\right) \left( \Phi _{2m}^{\left( 0,0\right) }+e^{-2\pi im/N}\Phi
_{2m}^{\left( 0,1/2\right) }\right) \\
&=&\frac{1}{\sqrt{2}}\Phi _{2m}^{\left( 0,0\right) }+\frac{e^{-2\pi im/N}}{%
\sqrt{2}}\left( E_{x}+X^{-1/2}O_{x}\right) \left( B-T\right) \Phi
_{2m}^{\left( 0,1/2\right) }.
\end{eqnarray*}
Now, observe that 
\begin{eqnarray*}
B\Phi _{m}^{(0,1/2)} &=&\frac{e^{i\pi m/N}}{\sqrt{N}}\sum_{k\in \mathbb{Z}%
}(-1)^{k}\int_{[0,1/2)+\mathbb{Z}}\left| p\right\rangle \left\langle
p\right| \left. \frac{m}{N}+k\right\rangle _{x}dp \\
&=&e^{i\pi m/N}\int_{[0,1/2)+\mathbb{Z}}\left| p\right\rangle \left(
\sum_{k\in \mathbb{Z}}e^{2\pi ik(1/2-Np)}\right) e^{-2\pi ipm}dp \\
&=&\frac{e^{i\pi m/N}}{\sqrt{N}}\sum_{a=0}^{N/2-1}e^{-2\pi i\left(
a+1/2\right) m/N}\;\frac{1}{\sqrt{N}}\sum_{k\in \mathbb{Z}}\left| \frac{a+1/2%
}{N}+k\right\rangle _{p} \\
&=&\frac{1}{\sqrt{N}}\sum_{a=0}^{N/2-1}e^{-2\pi iam/N}\;\widetilde{\Phi }%
_{a}^{(0,1/2)},
\end{eqnarray*}
where we have used the ``p-state'' $\delta $-comb given in \ref{p state
delta comb}. Thus, we see that 
\[
\left( B-T\right) \Phi _{m}^{(0,1/2)}=\frac{1}{\sqrt{N}}\left(
\sum_{a=0}^{N/2-1}e^{-2\pi iam/N}-\sum_{a=N/2}^{N-1}e^{-2\pi iam/N}\right)
\sum_{b=0}^{N-1}\left( \mathcal{F}^{-1}\right) _{ab}\;\Phi _{b}^{(0,1/2)}, 
\]
and 
\begin{eqnarray}
F\Phi _{m}^{\left( 0,0\right) } &=&\frac{1}{\sqrt{2}}\Phi _{2m}^{\left(
0,0\right) }  \label{actual sum} \\
&&+\frac{e^{-2\pi im/N}}{\sqrt{2N}}\left( \sum_{a=0}^{N/2-1}e^{-2\pi
ia\left( 2m\right) /N}-\sum_{a=N/2}^{N-1}e^{-2\pi ia\left( 2m\right)
/N}\right)  \nonumber \\
&&\times \sum_{b=0}^{N-1}\left( \mathcal{F}^{-1}\right) _{ab}\;\left(
E_{x}+\left( -1\right) ^{b}O_{x}\right) \Phi _{b}^{(0,1/2)}.  \nonumber
\end{eqnarray}
Now oberve 
\[
E_{x}\Phi _{m}^{(0,1/2)}=\frac{e^{i\pi m/N}}{\sqrt{N}}\sum_{k\in \mathbb{Z}%
}e^{i\pi k}\chi _{e_{x}}(\frac{m}{N}+k)\left| \frac{m}{N}+k\right\rangle
_{x}. 
\]
So for $m\in \left[ 1,N-1\right] $, this yields 
\begin{eqnarray*}
E_{x}\Phi _{m}^{(0,1/2)} &=&\frac{e^{i\pi m/N}}{\sqrt{N}}\sum_{k\in \mathbf{%
even}}e^{i\pi k}\left| \frac{m}{N}+k\right\rangle _{x}=\frac{e^{i\pi m/N}}{%
\sqrt{N}}\sum_{k\in \mathbb{Z}}\left| \frac{m}{N}+2k\right\rangle _{x} \\
&=&\frac{\Phi _{m}^{(0,1/2)}+e^{i\pi m/N}\Phi _{m}^{(0,0)}}{2}.
\end{eqnarray*}
We let $\left[ m/N\right] $ be the integer part of $m/N$. Then 
\begin{eqnarray*}
E_{x}\Phi _{m}^{(0,1/2)} &=&\frac{1}{2}\left( \Phi _{m}^{(0,1/2)}+e^{i\pi
\left( m/N-\left[ m/N\right] \right) }\Phi _{m}^{(0,0)}\right) , \\
O_{x}\Phi _{m}^{(0,1/2)} &=&\frac{1}{2}\left( \Phi _{m}^{(0,1/2)}-e^{i\pi
\left( m/N-\left[ m/N\right] \right) }\Phi _{m}^{(0,0)}\right) ,
\end{eqnarray*}
and hence 
\[
\left( E_{x}-O_{x}\right) \Phi _{m}^{(0,1/2)}=e^{i\pi \left( m/N-\left[
m/N\right] \right) }\Phi _{m}^{(0,0)}. 
\]
(Note that we have avoided the case $m=0$ since eqn. \ref{actual sum} has
only \textit{odd }terms in the sum. There is a subtelty involved in this
case coming from the fact that $\Phi _{m}^{(0,0)}$ occurs on the boundary.)

Thus, 
\begin{eqnarray*}
F\Phi _{m}^{\left( 0,0\right) } &=&\frac{1}{\sqrt{2}}\Phi _{2m}^{\left(
0,0\right) } \\
&&+\frac{e^{-2\pi im/N}}{\sqrt{2N}}\left( \sum_{a=0}^{N/2-1}e^{-2\pi
ia\left( 2m\right) /N}-\sum_{a=N/2}^{N-1}e^{-2\pi ia\left( 2m\right)
/N}\right) \sum_{b\,\text{even}}^{N-1}\left( \mathcal{F}^{-1}\right)
_{ab}\Phi _{b}^{(0,1/2)} \\
&&+\frac{e^{-2\pi im/N}}{\sqrt{2N}}\left( \sum_{a=0}^{N/2-1}e^{-2\pi
ia\left( 2m\right) /N}-\sum_{a=N/2}^{N-1}e^{-2\pi ia\left( 2m\right)
/N}\right) \sum_{b\,\text{odd}}^{N-1}e^{i\pi b/N}\left( \mathcal{F}%
^{-1}\right) _{ab}\;\Phi _{b}^{(0,0)}.
\end{eqnarray*}
Consider just the middle term. Since we have already shown that $\theta
=\left( 0,0\right) $ is a fixed point of the dynamics, we should see this
term exactly vanishing. In fact a direct calculation readily shows this. For 
$b$ even 
\begin{eqnarray*}
&&\frac{e^{-2\pi im/N}}{\sqrt{2N}}\left( \sum_{a=0}^{N/2-1}e^{-2\pi ia\left(
2m\right) /N}-\sum_{a=N/2}^{N-1}e^{-2\pi ia\left( 2m\right) /N}\right)
\left( \mathcal{F}^{-1}\right) _{ab}\; \\
&=&\frac{e^{-2\pi im/N}}{N\sqrt{2}}\left( \sum_{a=0}^{N/2-1}e^{-2\pi
ia\left( 2m\right) /N}e^{2\pi iab/N}-\sum_{a=0}^{N/2-1}e^{-2\pi i\left(
a+N/2\right) \left( 2m\right) /N}e^{2\pi i\left( a+N/2\right) b/N}\right) \\
&=&0.
\end{eqnarray*}
Thus, 
\begin{eqnarray*}
F\Phi _{m}^{\left( 0,0\right) } &=&\frac{1}{\sqrt{2}}\Phi _{2m}^{\left(
0,0\right) } \\
&&+\frac{e^{-2\pi im/N}}{\sqrt{2N}}\left( \sum_{a=0}^{N/2-1}e^{-2\pi
ia\left( 2m\right) /N}-\sum_{a=N/2}^{N-1}e^{-2\pi ia\left( 2m\right)
/N}\right) \sum_{b\,\text{odd}}^{N-1}e^{i\pi b/N}\left( \mathcal{F}%
^{-1}\right) _{ab}\;\Phi _{b}^{(0,0)}.
\end{eqnarray*}
We next calculate the matrix elements. We see 
\[
\left( \Phi _{n}^{\left( 0,0\right) },F\Phi _{m}^{\left( 0,0\right) }\right)
_{P}=\left\{ 
\begin{array}{ll}
\sum_{a=0}^{N/2-1}\left( \mathcal{F}^{N}\right) _{na}^{-1}\left( 
\begin{array}{ll}
\mathcal{F}^{N/2} & 0 \\ 
0 & 0
\end{array}
\right) _{am} & n\text{ even} \\ 
e^{i\pi \left( n-2m\right) /N}\sum_{a=0}^{N/2-1}\left( \mathcal{F}%
^{N}\right) _{na}^{-1}\left( 
\begin{array}{ll}
\mathcal{F}^{N/2} & 0 \\ 
0 & 0
\end{array}
\right) _{am} & n\,\text{odd}
\end{array}
\right. 
\]
For the case $N/2\leq m<N$, we see 
\begin{eqnarray*}
F\Phi _{m}^{\left( 0,0\right) } &=&\left( E_{x}+X^{-1/2}O_{x}\right) \left(
B+Y^{-1}T\right) \sqrt{\frac{2}{N}}\sum_{k\in \mathbb{Z}}\left| \frac{2m-N}{N%
}+2k+1\right\rangle \\
&=&\frac{1}{\sqrt{2}}\phi _{2m-N}^{\left( 0,0\right) }-\frac{e^{-2\pi im/N}}{%
\sqrt{2N}}\left( \sum_{a=N/2}^{N-1}e^{-2\pi ia\left( 2m\right)
/N}-\sum_{a=N/2}^{N-1}e^{-2\pi ia\left( 2m\right) /N}\right) \\
&&\times \sum_{b\,\text{odd}}^{N-1}e^{i\pi b/N}\left( \mathcal{F}%
^{-1}\right) _{ab}\;\Phi _{b}^{(0,1/2)}
\end{eqnarray*}
Thus, 
\[
\left( \Phi _{n}^{\left( 0,0\right) },F\Phi _{m}^{\left( 0,0\right) }\right)
_{P}=\left\{ 
\begin{array}{ll}
\sum_{a=0}^{N/2-1}\left( \mathcal{F}^{N}\right) _{na}^{-1}\left( 
\begin{array}{ll}
0 & 0 \\ 
0 & \mathcal{F}^{N/2}
\end{array}
\right) _{am} & n\text{ even} \\ 
-e^{i\pi \left( n-2m\right) /N}\sum_{a=0}^{N/2-1}\left( \mathcal{F}%
^{N}\right) _{na}^{-1}\left( 
\begin{array}{ll}
0 & 0 \\ 
0 & \mathcal{F}^{N/2}
\end{array}
\right) _{am} & n\,\text{odd}
\end{array}
\right. 
\]
Combining these results, we see that eqn. \ref{OURS} holds for any $m$ and $%
n $. This completes the proof of the theorem. 
%TCIMACRO{\TeXButton{End Proof}{\endproof}}
%BeginExpansion
\endproof%
%EndExpansion

It is easy to see that the matrix $\left( B\right) _{nm}=\left( \Phi
_{n}^{(0,0)},F\Phi _{m}^{(0,0)}\right) $ is unitary, since from eqn. \ref
{OURS} it is the product of four unitary $N\times N$ matrices.

\section{Symmetries in the Quantum Dynamics}

We conclude with a demonstration that the quantization prescription we use
preserves the two classical symmetries at $\theta =\left( 0,0\right) $: (1)
Parity, and (2) Time-Reversal.

(1) Parity

One of the benfits of the quantization prescription given above is the that
the classical parity operation $x\rightarrow 1-x$, $p\rightarrow 1-p$ is
preserved at the point $\theta =\left( 0,0\right) $ in the quantum dynamics.
Let us discuss the classical symmetry first. A map $\beta :\left( x,p\right)
\rightarrow \left( x^{\prime },p^{\prime }\right) $ is said to be symmetric
under the operation $s:\left( x,p\right) \rightarrow \left( \widetilde{x},%
\widetilde{p}\right) $ if $\beta ^{*}s^{*}f=s^{*}\beta ^{*}f$, where the $*$
denotes the pullback of the map on the classical algebra of observables
(here the periodic functions). It is easy to see that this property holds
classically. Acting on a harmonic $e^{2\pi i\left( ax+bp\right) }$, we see 
\begin{eqnarray*}
s^{*}\beta ^{*}e^{2\pi i\left( ax+bp\right) } &=&-e^{-4\pi iax}e^{-i\pi
bp}\left( \chi _{r}\left( x\right) +\left( -1\right) ^{b}\chi _{l}\left(
x\right) \right) \left( \chi _{o_{p}}\left( p\right) +\left( -1\right)
^{b}\chi _{e_{p}}\left( p\right) \right) \\
&=&-e^{-4\pi iax}e^{-i\pi bp}\left( \chi _{l}\left( x\right) +\left(
-1\right) ^{b}\chi _{r}\left( x\right) \right) \left( \chi _{e_{p}}\left(
p\right) +\left( -1\right) ^{b}\chi _{o_{p}}\left( p\right) \right) \\
&=&\beta ^{*}s^{*}e^{2\pi i\left( ax+bp\right) }.
\end{eqnarray*}
The case of a more general function follows readily from linearity and
continuity.

In the quantum dynamics, we look only at the case $\theta =\left( 0,0\right) 
$ and we define the operation of conjugation as 
\begin{eqnarray*}
P\left| x\right\rangle &=&\left| -x\right\rangle , \\
P\left| p\right\rangle &=&\left| -p\right\rangle
\end{eqnarray*}
Observe that $P=P^{\dagger }=P^{-1}$and 
\[
PU^{a}V^{b}P=U^{-a}V^{-b}. 
\]
In fact it is easy to see that on the Hilbert space $\mathcal{H}_{\hbar
}\left( 0\right) $ is invariant under $P$. Explicitly, 
\[
P\Phi _{m}^{\left( 0,0\right) }=\Phi _{N-m}^{\left( 0,0\right) } 
\]
and 
\begin{eqnarray*}
P\widetilde{\Phi }_{n}^{\left( 0,0\right) } &=&\frac{1}{\sqrt{N}}%
\sum_{m=0}^{N-1}e^{-2\pi inm/N}P\Phi _{m}^{\left( 0,0\right) } \\
&=&\frac{1}{\sqrt{N}}\sum_{m=0}^{N-1}e^{-2\pi inm/N}\Phi _{N-m}^{\left(
0,0\right) } \\
&=&\frac{1}{\sqrt{N}}\sum_{m^{\prime }=0}^{N-1}e^{-2\pi i\left( N-n\right)
m/N}\Phi _{m}^{\left( 0,0\right) } \\
&=&\widetilde{\Phi }_{N-n}^{\left( 0,0\right) }.
\end{eqnarray*}

It is now straightforward to check that $PFP=F$ on $\Phi _{m}^{\left(
0,0\right) }$. Consider first the case $n>0$, even. Then 
\begin{eqnarray*}
\left( PFP\right) _{nm} &=&\frac{1}{\sqrt{2}}\left( \delta
_{N-n,2N-2m}+\delta _{N-n,N-2m}\right) \\
&=&\frac{1}{\sqrt{2}}\left( \delta _{n,2m}+\delta _{n,2m-N}\right) =F_{nm}%
\text{.}
\end{eqnarray*}
At $n=0$, 
\[
F_{0m}=\frac{1}{\sqrt{2}}\left( \delta _{0,m}+\delta _{N/2,m}\right) 
\]
but $P$ is simply the identity operator on $\Phi _{0}^{\left( 0,0\right) }$%
and $\Phi _{N/2}^{\left( 0,0\right) }$.

For the case $n$ odd, we simply multiply the Balazs-Voros matrix elements by
the extra phase $\exp \left( i\pi (\left( n-2m\right) /N-\left[ 2m/N\right]
)\right) $to get 
\[
F_{nm}=\frac{e^{i\pi \left( n-2m\right) /N}}{N\sqrt{2}}\left( 1+i\cot \left(
\pi \left( n-2m\right) /N\right) \right) \text{.} 
\]
Thus 
\begin{eqnarray*}
\left( PFP\right) _{nm} &=&-\frac{e^{i\pi \left( 2m-n\right) /N}}{N\sqrt{2}}%
\left( 1+i\cot \left( \pi \left( 2m-n\right) /N\right) \right) \\
&=&-\frac{e^{i\pi \left( 2m-n\right) /N}}{N\sqrt{2}}\left( 1-\frac{e^{i\pi
\left( 2m-n\right) /N}+e^{-i\pi \left( 2m-n\right) /N}}{e^{i\pi \left(
2m-n\right) /N}-e^{-i\pi \left( 2m-n\right) /N}}\right) \\
&=&-\frac{e^{i\pi \left( 2m-n\right) /N}}{N\sqrt{2}}\left( \frac{-2e^{-i\pi
\left( 2m-n\right) /N}}{e^{i\pi \left( 2m-n\right) /N}-e^{-i\pi \left(
2m-n\right) /N}}\right) \\
&=&F_{nm}\text{.}
\end{eqnarray*}
Thus on the subspace $\mathcal{H}_{\hbar }\left( 0\right) $ parity is
conserved. Observe that the matrix propagator we get is slightly different
from that given in \cite{BV}. We see now that the extra phases (which vanish
as $\hbar \rightarrow 0$) are precisely the terms necessary to return this
classical symmetry in the quantum dynamics.

(2) Time-Reversal

The classical baker's map also exhibits a time-reversal symmetry under the
following transformation: 
\begin{eqnarray*}
x &\rightarrow &p, \\
p &\rightarrow &x, \\
t &\rightarrow &-t\text{.}
\end{eqnarray*}
The quantization given in ref. \cite{BV} exhibits this symmetry, and, as we
shall now verify, the quantum propagator we have given here does as well. We
define an anti-linear operator $\Omega $ which takes $x$ to $p$ through its
action on a position eigenstate $\left| a\right\rangle _{x}$: 
\[
\Omega \left| a\right\rangle _{x}=\left| a\right\rangle _{p}\text{.} 
\]
In other words, $\Omega $ transforms a state at position $a$ to a state with
momentum $a$. We can now easily see that $\Omega $ has a similar action on a
momentum eigenstate: 
\begin{eqnarray*}
\Omega \left| a\right\rangle _{p} &=&\Omega \int e^{ixa/\hbar }\left|
x\right\rangle _{x}\frac{dx}{\sqrt{2\pi \hbar }} \\
&=&\int e^{-ixa/\hbar }\left| x\right\rangle _{p}\frac{dx}{\sqrt{2\pi \hbar }%
} \\
&=&\left| a\right\rangle _{x}
\end{eqnarray*}
where in the middle step we use the anti-linearity of $\Omega $. Indeed,
from this it follows that $\Omega ^{-1}=\Omega $ and 
\begin{eqnarray*}
\Omega \widehat{x}\Omega &=&\widehat{p}, \\
\Omega \widehat{p}\Omega &=&\widehat{x}.
\end{eqnarray*}
We see immediately the action on the generators of the quantum torus: 
\begin{eqnarray*}
\Omega U\Omega &=&V^{-1}, \\
\Omega V\Omega &=&U^{-1}.
\end{eqnarray*}
We can now readily calculate the action of $\Omega $ on different components
of the propagator. For example, 
\begin{eqnarray*}
\Omega E_{x}\Omega \left| a\right\rangle _{p} &=&\Omega \int_{[0,1)+2%
\mathbb{Z}}\left| x\right\rangle _{x\,x}\left\langle x\right| \left.
a\right\rangle _{x}dx \\
&=&\int_{[0,1)+2\mathbb{Z}}\delta \left( x-a\right) \left| x\right\rangle
_{p}dx \\
&=&E_{p}\left| a\right\rangle _{p}\text{.}
\end{eqnarray*}
This last equality is due to the fact that $x$ is really just a dummy
variable being integrated over. Similarly, we find 
\[
\Omega O_{x}\Omega =O_{p},\quad \Omega L\Omega =B,\quad \Omega R\Omega =T.%
\text{\quad } 
\]
Also, since $\Omega $ takes $\widehat{x}\rightarrow \widehat{p}$, $%
i\rightarrow -i$, we find 
\[
\Omega X^{-1/2}\Omega =Y^{1/2}\quad \Omega Y^{-1}\Omega =X,\text{\quad }%
\Omega S\Omega =S^{\dagger }\text{.} 
\]
Thus on the propagator (eqn. \ref{propagator}), 
\begin{eqnarray*}
\Omega F\Omega &=&\Omega \left( E_{x}+X^{-1/2}O_{x}\right) \left(
B+Y^{-1}T\right) S\Omega \\
&=&\left( E_{p}+Y^{1/2}O_{p}\right) \left( L+XR\right) S^{\dagger } \\
&=&S^{\dagger }\left( B+YT\right) \left( E_{x}+X^{1/2}O_{x}\right) \\
&=&F^{\dagger }.
\end{eqnarray*}
True to its classical origin, the full quantum propagator exhibits the
appropriate time-reversal symmetry. What this means is that time-reversal
symmetry holds \textit{at all values on the }$\theta $\textit{-torus. }We
can see easily how to implement time reversal at the point $\theta =\left(
0,0\right) $: 
\begin{eqnarray*}
\Omega \Phi _{m}^{\left( 0,0\right) } &=&\Omega \left( \frac{1}{\sqrt{N}}%
\sum_{k}\left| \frac{m}{N}+k\right\rangle _{x}\right) \\
&=&\frac{1}{\sqrt{N}}\sum_{k}\left| \frac{m}{N}+k\right\rangle _{p} \\
&=&\widetilde{\Phi }_{m}^{\left( 0,0\right) }=\sum_{n=0}^{N-1}\left( 
\mathcal{F}^{N}\right) _{mn}^{-1}\Phi _{n}^{\left( 0,0\right) }\text{.}
\end{eqnarray*}
Thus on the subspace $\mathcal{H}_{\hbar }\left( 0\right) $, $\Omega $ can
be implemented via a matrix Fourier transform combined with complex
conjugation. This is precisely the form used in ref. \cite{BV} to
demonstrate time-reversal symmetry for the Balazs-Voros matrices.

\section{Acknowledgements}

The authors would like to thank Andrew Lesniewski, Christopher King, Lev
Kaplan, Eric Heller, Jon Tyson and Sidney Coleman for many fruitful
discussions. The authors would also like to thank the referee for insightful
and constructive comments which we hope have made the paper more readable.

\smallskip \pagebreak \FRAME{ftbpFU}{224pt}{97.25pt}{0pt}{\Qcb{The baker's
map on the torus. The square gets squished to half its height and stretched
to twice its length, and the right region gets chopped off and placed back
on top.}}{}{Figure 1}{\special{language "Scientific Word";type
"GRAPHIC";maintain-aspect-ratio TRUE;display "USEDEF";valid_file "T";width
224pt;height 97.25pt;depth 0pt;original-width 429.625pt;original-height
185pt;cropleft "0";croptop "1";cropright "1";cropbottom "0";tempfilename
'C:/SWP25SE/docs/EW94KE00.wmf';tempfile-properties "XP";}}

\smallskip

\smallskip

\smallskip

\FRAME{fphFU}{296.3125pt}{108.5625pt}{0pt}{\Qcb{One iteration of the
classical baker covering map. Observe that all ``LEFT'' regions map to
``BOTTOM'' regions, and ``TOP'' to ``RIGHT''. In this way, acting on
periodic functions the covering map is exactly the baker's map on the torus.}%
}{}{Figure 2}{\special{language "Scientific Word";type
"GRAPHIC";maintain-aspect-ratio TRUE;display "USEDEF";valid_file "T";width
296.3125pt;height 108.5625pt;depth 0pt;original-width
504.0625pt;original-height 182.8125pt;cropleft "0";croptop "1";cropright
"1";cropbottom "0";tempfilename
'C:/SWP25SE/docs/EW94KE01.wmf';tempfile-properties "XP";}}

\pagebreak 
\[
\begin{tabular}{ccc}
& Table 1 &  \\ 
&  &  \\ 
Operator $\left( \mathcal{O}\right) $ & $\left( x_{0},p_{0}\in ?\right) $ & $%
\lim_{\hbar \rightarrow 0}\left\| \mathcal{O}\left| \phi _{\hbar
};x_{0},p_{0}\right\rangle \right\| $ \\ 
$E_{x}$ & $\tilde{e}_{x}$ & $1$ \\ 
$E_{x}$ & $\tilde{e}_{x}$ & $0$ \\ 
$O_{x}$ & $\tilde{o}_{x}$ & $0$ \\ 
$O_{x}$ & $\tilde{o}_{x}$ & $1$ \\ 
$E_{p}$ & $\tilde{e}_{p}$ & $1$ \\ 
$E_{p}$ & $\tilde{e}_{p}$ & $0$ \\ 
$O_{p}$ & $\tilde{o}_{p}$ & $0$ \\ 
$O_{p}$ & $\tilde{o}_{p}$ & $1$ \\ 
$L$ & $\widetilde{l}$ & $1$ \\ 
$L$ & $\widetilde{l}$ & $0$ \\ 
$R$ & $\widetilde{r}$ & $0$ \\ 
$R$ & $\widetilde{r}$ & $1$ \\ 
$B$ & $\widetilde{b}$ & $1$ \\ 
$B$ & $\widetilde{b}$ & $0$ \\ 
$T$ & $\widetilde{t}$ & $0$ \\ 
$T$ & $\widetilde{t}$ & $1$%
\end{tabular}
\]

\end{document}